# Integrating Vague Association Mining with Markov Model

Priya Bajaj[1] and Supriya Raheja[2]

[1,2]Department of Computer Science & Engineering, ITM University
Gurgaon, Haryana 122001, India

**ABSTRACT**

*The increasing demand of World Wide Web raises the need of predicting the user's web page request. The most widely used approach to predict the web pages is the pattern discovery process of Web usage mining. This process involves inevitability of many techniques like Markov model, association rules and clustering. Fuzzy theory with different techniques has been introduced for the better results. Our focus is on Markov models. This paper is introducing the vague Rules with Markov models for more accuracy using the vague set theory.*

## 1. Introduction

The World Wide Web (WWW) has turned to an immense, variant, and dynamic information reservoir accessed by people with different backgrounds and interests. As number of internet users has grown a lot now a day which has ultimately resulted increase in user's web access latency. If the latency is high then the web page requested by user takes more time which leads poor quality of service. To resolve this problem Web usage mining is used in which the user behaviour is discovered and the upcoming web pages are predicted, then pre-fetched and cached in advance to minimize the response time of retrieving the web document. Predicting the user's next requested web page can be achieved through information collected by Web server's logs. .- Web mining and user modelling are those techniques that make use of these access data, discover the browsing patterns, and amend the efficiency of Web surfing. Web Log is a rapid and powerful access log analyser. It provides user's information such as : Internet Protocol address of the system, approached files and date and time of accessing those files, action performed by user, paths from the site, information of referred pages, information about used search engines and browsers, operating systems on which user is operating, and more[11]. User's navigation sessions are prepared with the help of Web log file [20]. Several models are available to model user navigation sessions such as statistical analysis, association rules and classification. But the most widely used technique for preparing the web navigation sessions is Markov model which is based on a well-proved theory and easy to understand. Markov model executes with consecutive sequential approach of web pages then the recently approached requests which have more impact on user's next request is considered in the prediction process[2]. Markov model is of two type's namely higher order and lower order. Lower order Markov models covers limited browsing history of user and therefore results in low accuracy. Higher order Markov models results in high accuracy but are associated with higher state space complexity. [11] The sequential coverage results in higher precision because the click stream data is inherently sparse and Markov models does not work well with the sparse data . Using above techniques prediction is made and then according to that prediction web pages are pre-fetched and placed in web caches. So when a user asked for a web page, that page is first looked in web cache first. Caching is a temporary memory space where web page data is stored. Web caching is similar to caching in memory system .A Web cache contains Web resources in expectation of future requests. Caching works out on various locations over the Web, including at the two end-points i.e. the Web browser and Web server. Web caching is based on popularity- the more popular a resource is, the more priority it is to be requested in the future. Advantages of caching are [2, 3]:-

(1)Web Caching reduces the client access latency.

(2) By decreasing the network traffic and reducing the network congestion, web caching reduces the bandwidth consumption

(3) Web caching reduces the workload of remote Web server by circulating data among the caches over the wide area network.

(4) Caching enhances the robustness of the remote server. If server cannot send the response or there is some other problem with network then the data can directly be accessed from caching.
When the request made by user is satisfied by the cache , the content have no longer need to travel across the Internet from the root web server to the cache, saving bandwidth be cached too long [2]. In this paper, we propose different techniques of web page prediction and study use of fuzzy and vague based approach for discovering users' navigational paths. In particular, we are interested in predicting a user's next re-quest as well as the time that user is likely to spend on that request. When the prediction is only based on simple time categories without fuzzification, the problem of sharp boundaries may affect the accuracy. Therefore, our proposed method merges the fuzzy theory to address this problem with the aim of improving the prediction accuracy.

## 2. Related Work

A number of experiments have been made by researchers to improve the Web page access prediction precision by combining different recommendation frameworks.

R.R. Sarukkai [4] deals with first-order Markov models to form the sequence of pages requested by a user to predict the next page accessed. An "Alterd" Markov model is trained for every particular user and used for predicting the user's future requests. In practice, it is very expensive to construct a unique model for every user respectively, and the situation gets even worse when there exist thousands of different users within a big Web site.

V. Padbanabham and J. Mogul[5] deals with N-hop Markov models for predicting the next web page that user accesses most very frequently by P-matching the user's current session with the user's historical web sessions for improving pre-fetching strategies for web caches.

F. Khalil [6] presents the Integration Prediction Model (IPM) by merging Association rules, clustering algorithm and Markov model together. Then, the cluster sets are used to perform prediction rather than the actual sessions. There are different constraint on which the IPM integration model is based. The web user sessions initially are divided into variety of separate clusters by applying k-means clustering algorithm and cosine distance measure. Then, the integration model estimates Markov model divination on the concluding clusters. This will amend the state space complexity because Markov model prediction is exercised on the specific clusters as adverse to the whole data set. Association rules are examined on more states than Markov model by taking more history if the state is absent in the training dataset. In the end, if a new page is introduced, the cosine distance is computed and identifies an appropriate cluster to which a new web page should belong. The integration model has been proved by the experiments which improve the prediction accuracy. Moreover, applying the prediction model on the clusters gives better results than on the non-clustered data. Even if/though, the web page access prediction result was improved, however, it can be but we can see that their integrated algorithm has complicated procedures and must repeatedly apply in order to increase their prediction performance.

S. Chimphlee[7] proposed a Hybrid Markov Fuzzy Models (HyMFM) which are obtained by merging the benefits of all the three prediction models: Association rules, Markov model and Fuzzy Adaptive Resonance Theory (Fuzzy ART). Hybrid Markov Fuzzy Model algorithm clusters the web user sessions by presenting the new sequence and the new similarity measures in augmentation learning of Fuzzy ART control structure. An internet user session was displayed into the transition matrix representation, ascribed as session matrix that is prepared on the basis of transition matrix of a leading Hybrid Markov model. Both elements are suitable for the clustering task that is used to treat the web user sessions as order sets of accesses. Consequently, to enable the application of Fuzzy ART clustering the new similarity measures were evolved. This study states two new similarity measures: Matrix distance similarity and Matrix norm similarity. These similarity measures mitigate the overvalue problem in Fuzzy ART rule which utilizes the city-block distance metric because the similarity measure between prototypes and inputs. Thus, the web user sessions were assembled into clusters with similar patterns during the training phase and when it brings in the front of the new input, it produces the result that indicates cluster to which it belongs and then Hybrid Markov model is executed on each cluster.

## 3. Fuzzy Set

Zadeh [8] proposed the theory of fuzzy set for handling imprecise decision-making problems. In fuzzy set theory, each object o ∈ U is assigned a single real value, which is known as the grade of membership, between zero to one.

For example A Fuzzy Set S in X is characterized by a membership function $f_s(Z)$ ,where each point in Z is a real number in the interval [0, 1] with the value of $f_s(Z)$ at x exhibiting the grade of membership of Z in fuzzy set S .Basically, Fuzzy Logic i.e. a multi valued or many valued logic, that supports intermediate values to be defined as conventional evaluations like high /low, fast/slow, true/false, hot/cold etc.

### 3.1 Fuzzy Mining Approach for Web Page Prediction

Mining association rules is a widely used method for qualitative data mining. It has been improved by dividing quantitative attributes into regions (categories). To address the problem of sharp boundaries, Gyenesei proposed a fuzzy mining method in[9]. As we know session is made of a time-oriented sequence of requests, association rules are not suitable for the direct prediction of a user's next request. Fuzzy logic can also use for the purpose of prediction. Data extracted form past record or history provides the information about the specific pattern of data learning and by the past learning, knowledge about the future to some extent can be gained ,fuzzy logic prediction system works basically on this concept. It basically carries two steps clustering and specification of input and output relations by using IF-THEN rules. FL provides a simple way which enables us to reach at the definite conclusion which is based upon vague, imprecise, ambiguous noisy, or missing input information.

A fuzzy membership function [10] signifies the likelihood of a value related to a fuzzy set. Generally, this kind of function maps the attribute values to a real number interval [0, 1], $\mu A : X \rightarrow [0, 1]$, where μ is the membership function, A is the fuzzy set, and X is the domain. For convenience, the fuzzy membership function is represented by $\mu_F$, where the subscript f denotes the corresponding fuzzy set. The total number of fuzzy sets is represented by S (i.e., $f \in [1, S]$). If we set S = 3, then f = 1, 2 and 3 can imply the fuzzy sets short, medium, and long, respectively. This represents the attributes of time-duration. By analyzing the time-duration for which the Web pages are visited, these three corresponding functions are derived. In particular, there may exist some special cases, suppose $\mu_1 (0) = 1.0$ means zero second has a full degree (1.0) on the "short" function (i.e., zero time-duration is considered as entirely short). On the other hand, if we take a relatively large number MAX TD as the time-duration threshold, we have

$$\mu_f (\tau) = 1.0, \forall \tau > MAXTD$$

For each time-duration $\tau$, $\mu_f (\tau) \in [0,1]$, theoretically, the sum of the values for a particular is between 0 and S. However, is associated to one request and therefore should only contribute to the statistics once per calculation. It is unreasonable for this sum to be greater or less than 1 (it has to be exactly 1). Thus, we need to normalize the functions.

After normalization, we have

$$\sum_{f=1}^{S} \bar{\mu}_f (\tau) = 1.0, \forall \tau \in [0, MAX\ TD],$$

Where $\bar{\mu}_f$ is the normalized membership function of the fuzzy set f.

Prediction models that already exist extract past navigation pattern of user to predict their next action. Past navigational information is gathered from Web log files, cookies and forms used by customers. Mainly two types of approaches are used for prediction: point-based approach, which uses frequency measures to perform prediction about user's next requests; and path-based approach, uses statistical analysis of user's previous navigation path. It is a complex task for we usage mining to identify navigation pattern. Usually Mining association rules are used for qualitative data mining. It has been amended by assigning quantitative attributes into regions. Classical association rules do not consider the time order of the attributes. Therefore association rules are not suitable for the direct prediction of a user's next request. So it is apparent that predicting the sequences of the user's next actions is a topic that has been examined from many angles, and various methods have been exercised in an attempt to find the best solution. We have tried to provide a contribution to this challenging topic by taking a slightly different approach.

## 4. Proposed Work

With the popularity of the internet the Web Mining becomes one of the major research areas. The web mining is further categorized as the Web Uses Mining and Web Content Mining. We are focusing in the same area to track the user activities on the internet by applying the Web Uses Mining. This approach is based on the concept of excel the web access regarding pre-fetching of the webpage on the basis of web page prediction. The proposed work is about to define effective approach for web page prediction. In this present work, the integration of Markov model is been defined with Vague based rule generation. The work is about to estimate the next visiting page effectively. At the initial layer, the Markov model is used to identify the paired frequency of web pages. Once the frequency analysis of user page visit is done, the value based rule set will be implemented to perform the frequency based classification of web pages. Finally the highest frequency page pair will be presented as the final result.

## 5. Research Methodology

Web page prediction includes pre-fetching of the next page to be accessed by the user or the link on which the Web user will click next while browsing a Web site. For an instance, what is the chance that a Web user visiting a site that sells computers will buy an extra mouse when buying a laptop? Or, there may be a greater chance the user will buy an external floppy drive instead. User's past browsing pattern is very radical concept in deriving such information. To perform this, each approach requires the browsing information in the form of server log file. Once browsing information is collected, the next work is to identify the user activity over the internet. By performing this analysis, user current visit is been identified along with identifying the next possible pages that will be visited by the user. The most visited page in combination in previous history will be cached in memory while browsing the current web page. The basic model of this pre-fetching process is shown in figure 1.

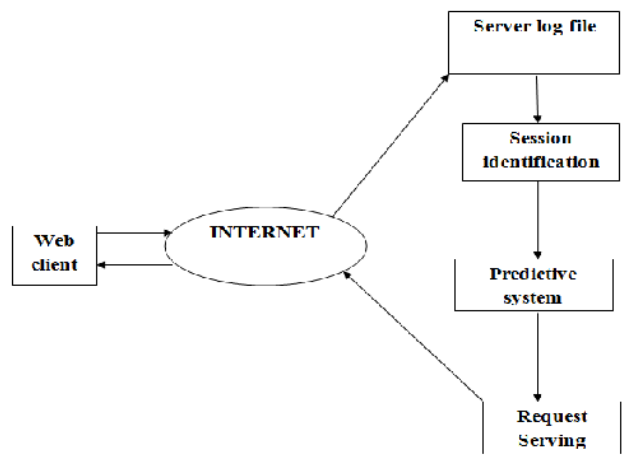

Figure.1 Pre-fetching Process

### 5.1 Server log file

Server log is a log file which automatically get created and maintained by a server of activity performed by it. Any request that is made to the Web server is displayed in this file. When a client requests for a web page to server, server stores that request as an entry to server log. The request recorded in this log file includes various kinds of activities, including the addressing of Internet Protocol of the computer cum making of systems cum creating or producing the desired request, date and time access, the desired document or image and so on.

### 5.2 Session Identification

A single user can make a group of requests, this group of requests is called a session for a single navigation purpose. A user can have a single session or multiple sessions during a single period of time. Here session is identified by taking that request from server log and those sessions are used in the predictive system.

### 5.3 Predictive system

The predictive system creates a graph using those identified sessions .Where each node is treated as a web page containing some other attributes. And an edge exhibits the link to go from one page to another. And after that it maintains the graph with updates by calculating the probability and producing new edges and nodes.

### 5.4 Request serving

The predicted pages are pre-fetched and maintain it in server cache and satisfies the user's request which diminishes the accessing time of that page and enhances the performance of web server.

## 6. Prediction Techniques

### 6.1 Association Rules

Major technique for discovering pattern is Association rules [11] Rather than considering the sequence of pages association rules uses the relationship of co-occurrence of pages. Due to because of this Association rules usually produce low exactness or precision, but high recall in prediction. An association rule X Y exhibits a relationship between the sets of items X and Y. Each item A is an atom representing an individual object. A direct association rule is the relationship X Y, where $X \subseteq D$, $Y \subseteq D$ and X Y = $\emptyset$. A direct association rule can be explained by two factors: support and confidence. The direct association rule X Y has the support. Support of the rule (X Y) is the percentage of transaction that contain both X and Y. The counter is increased by one every time for that web page, when the page is visited in different session [7]. Confidence of the rule is defined as percentage of number of session that contain X and y to the total no. of records that contain X where if some percentage exceeds the threshold of confidence an attractive and interesting rule can be generated. Association rule generation is used to describe pages that are most often referred together in single session. In the context of web usage mining, association rules ascribe those set of pages that are accessed together with a support value exceeding some specified threshold .The association rule may also serves as heuristic for pre-fetching document in order to reduce user perceived latency when loading a page from remote site .The rules are used in order to reveal correlation between pages accessed together during a server session. Such rule signifies the possible relationship between pages that are often visited together even they are not connected with each other directly and can discover association between groups of user with specific interests. The page set X is the body (or antecedent)and Y is the head (or consequent)of the rule X Y .The main restriction of association rules is that they produce many r rules and it is hard to find an appropriate subset of rules to make accurate reliable prediction.

### 6.2 Clustering

The most important unsupervised learning technique i.e clustering is defined as "the process of assembling objects into groups whose members in certain way have some similarities to one other "[12]. It means that a cluster is a assemblage of objects which are "homogeneous" and are "heterogeneous" to the objects which are not belonging to that clusters .The significant goal of clustering is to decide the intrinsic grouping in a set of unlabelled data. The important component of a clustering algorithm is measuring the distance between data points. Clustering of non-sequential data decreases the accuracy of traditional recommendation model

### 6.3 Markov Model

Markov models [13,14,15,16,17] have been extensively used for predicting the page that user will visit from the given pages which he or she has already visited. Markov models are iterative (recursive) decision trees that are utilized for modelling conditions that have events that may occur recursively over time or for modelling predictable events that occur over time. . The input data for constructing Markov models is composed of web-sessions, where each session contains the sequence of the pages accessed by the user during his/her visit to the site. This technique gives more precise result by considering the consecutive orders of preceding pages. Input for this problem will be sequence of HTML pages or web pages that were accessed by user and it is supposed that it is following Markov property.

$$P_a(x) = P_a(x_n | x_{n-1}) P_a(x_{n-1} | x_{n-2}) \ldots P_a(x_2 | x_1) = P_a(x_1) \tag{3}$$

$$\sum_{n=2}^{n} P_a(x_n | x_{n-1}) \tag{4}$$

This shows that the probability of a state will be based on probability of earlier visited pages but additional memory can be obtained in to states with the help of higher order Markov model. The simplest Markov model considers only the last action performed by the user for predicting the next action. Higher order Markov model will consider more history then the lower order Markov models. Lower order Markov models are less accurate, whereas higher order Markov models exhibits high state-space complexity and also the sequences are not present in Web-log. This motivates us to integrate some other feature which should be considered with lower order Markov Model for better prediction accuracy.

## 7. Vague Set

Vague sets are describd by Gau and Buehrer[18] which have an additional potential edge over fuzzy sets of Zadeh[8]. Vague set is a collection of objects, each set features a grade-of membership whose utility is a continuous subinterval of [0,1]. These sets are characterized by a truth-membership function and a false membership function. The selection of the membership boundary also has interesting implications on modelling relationship between vague data .The true membership value $t_R$ (x, y) computes the strength of the subsisting the relation of R-type of the item x with the item y, since the false membership value $f_R$ (x, y) computes the intensity of the non-existence of the relation of R-type of the item x with the item y.

Vague sets are also entitled to be called as higher order fuzzy set. A vague set A in V is by a truth-membership function $t_A$ and a false-membership function $f_A$,

$$t_A : V \to [0,1] \quad , \quad f_A : V \to [0,1] \tag{5}$$

where $t_A(u_i)$ is a lower bound on the grade of membership of $u_i$ is obtained from the evidence for $u_i$; $f_A(u_i)$ is a lower bound on the negation of $u_i$ obtained from the evidence against $u_i$, and $t_A(ui) + f_A(ui) \leq 1$. The grade of membership of $u_i$ in the vague set A is confined to a subinterval $[t_A(ui); 1 - f_A(ui)]$ of [0,1]. The vague value $[t_A(ui); 1 - f_A(ui)]$ signifies that the exact grade of membership $\mu_A(u_i)$ of $u_i$ may be unknown. But it is restricted by

$$t_A(ui) \leq \mu_A(ui) \leq 1 - f_A(ui), where\ t_A(ui) + f_A(ui) \leq 1 \tag{6}$$

Vague Sets are based on an interval based membership and thus more expressive in capturing vagueness of data. Therefore this kind of reasoning is also known as interval membership, as adverse to point membership in the context of fuzzy sets.

## 8. Vague Association Rules

The main feature of this approach is that this theory reveals the drawback of in fuzzy set theory that it uses single membership value which unfortunately does not allow a separation of evidence for membership and evidence against membership by using interval-based membership that apprehend three types of evidence with respect to an object in a universe of discourse: support, against and hesitation [22,23]. So, we can easily model the hesitation information of an item for the mining context as the evidence of hesitation with respect to the item. To study the relationship between the support evidence and the hesitation evidence with respect to an item, the concepts of attractiveness and hesitation of an item was introduced, which are based on the median membership and the imprecision membership that are derived from the vague membership in vague sets.

## 9. Conclusion

User's navigation path is an important aspect of user's behaviour. Predicting users navigation paths helps us to better grasp the characteristics of user surfing. Web page pre-fetching has been widely used to reduce the access latency problem of the internet; its result mainly relies on the accuracy of web page prediction. The vague idea can be used in quantitative data mining like Speed of a vehicle is fast or slow is a variable whose value can lie in probable range defined by quantitative parameters or limits. Our method incorporates vague logic into Markov chain in order to predict the page requests. Our method Vague Markov model grasped the characteristics of user surfing in a more accurate way.

## 10. Future Work

In future we will implement Markov model with vague association rules using Matlab.

## References

[1] Brala Meenu et al, "An Improved Markov Model Approach to Predict Web Page Caching," International Journal of Computer Science & Communication Networks, Volume No. 2(3), pp: (393-399)
[2] Davison B.D ,July 2001, "A Web Caching Primer," Volume No. 5 Issue 4, IEEE Educational Activities Department Piscataway, NJ, USA, pp: (38-45)
[3] Liu Yanjun, Wang Yueping and Du Huilin, August 2010, "Strong Cache Consistency on World Wide Web," 3rd International Conference on Advanced Computer Theory and Engineering(ICACTE),volume 5, pp: (62-62)
[4] Sarukkai, R. (2000), `Link prediction and path analy-sis using markov chains', 9th International WWW Conference, Amsterdam pp:(377-386).
[5] Padmanabhan V.N, Mogul J.C., 1996 "Using Predictive Prefetching to Improve World Wide Web Latency," Computer Communication Review
[6] Khalil, F , J. Li and H. wang ,2008, "Integrating recommendation models for improved web pages prediction accuracy" Proceedings of the 31th Australasian Computer Science Conference,(ACSC'08), Wollongong, NSW ,pp :(91-100)
[7] Chimphlee.S, Salim.N, Ngadiman M.S.B., 2010, "Hybrid Web Page Prediction Model for Predicting a User 's Next Access," Information Technology Journal, pp:(774-781).
[8] Zadeh L.A, 1965, "Fuzzy sets" Informaton and Control 8, pp: (338-353)
[9] Gyenesei, 2000, "A Fuzzy Approach for Mining Quantitative Association Rules," Turku Centre for Computer Science, TUCS Technical Reports, Turku, Finland, no. 336,.
[10] Ghorbani Ali A. and Xu Xiaowen,2007 , "A Fuzzy Markov Model Approach for Predicting User Navigation," IEEE/WIC/ACM Intenational Conference on Web Intelligence

[11] Kotsiantis Sotiris, Kanellopoulos Dimitris, 2006, "Association Rules Mining: A Recent Overview," International Transactions on Computer Science and Engineering, Vol.32 (1)

[12] Kumar Sunil, Kalra Mala,2013, "Web Page Prediction Techniques: A Review," (IJCTT) – volume 4 Issue 7,pp: (2062-2066)

[13] Nigamand Bhawna, Dr. Jain Suresh 2010, "Analysis Of Markov Model On Different Web Pre-fetching And Caching Schemes," IEEE

[14] Deshpande, M. & Karypis, G. ,2004, "Selective Markov models for predicting web page accesses," Transactions on Internet Technology 4 (2), pp: (163-184).

[15] Maheswara V.V.R, Rao & Dr. Kumari V.Valli, "An Efficient Hybrid Successive Markov Model for Predicting Web User Usage Behavior using Web Usage Mining," International Journal of Data Engineering (IJDE) Volume (1): Issue (5),pp: (43-62)

[16] Khanchana.R ,Dr. Punithavalli. M, 2011, "An Efficient Web Page Prediction Based on Access Time-Length and Frequency," IEEE

[17] Dubey Shreya, Mishra Nishchol May 2011 "Web Page Prediction using Hybrid Model," International Journal on Computer Science and Engineering (IJCSE), Volume. 3 No. 5,pp: (2170-2176)

[18] Gau Wen-Lung and Buehrer Daniel , March April 1993 "Vague Sets," IEEE Transactions On Systems, Man, & Cybernetics, VOL. 23, NO. 2, pp: (610-614)

[19] Khan Hakimuddin, Ahmad Musheer and Biswas Ranjit ,MARCH 2007 "Vague Relations," International Journal Of Computational Cognition VOL. 5, NO. 1, pp: (31-35)

[20] Lu An and Ng Wilfred , 2005 "Vague Sets or Intuitionistic Fuzzy Sets for Handling Vague Data: Which One Is Better?," Springer Berlin Heidelberg

[21] Jiehao Yu, Yang Cao, Yan Quan Shuang, Ping Jian, 2008 , "Distance Measure Between Vague Sets," IEEE

[22] Lu An, Ke Yiping, Cheng James, and Ng Wilfred, 2007,"Mining Vague Association Rules," Springer-Verlag Berlin, Heidelberg, 12th International Conference On DataBase System foe Advance Application pp: (891-897)

[23] Lu An, Ng Wilfred , 2007, " Mining Hesitation Information by Vague Association Rules," Springer-Verlag Berlin, Heidelberg 26th international conference on conceptual modeling, pp:(39-55)

**Authors**

First Author **Priya Bajaj**, Have completed engineering in Computer Science from Maharishi Dayanand University, Rohtak in 2012 and pursuing M-tech in Computer Science from ITM University (2012-2014).

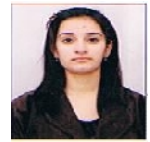

**Second Author Supriya Raheja, Assistant Professor,** ITM University, is pursuing her PhD in Computer Science from Banasthali University. She had done her engineering from Hindu college of Engineering, Sonepat and masters from Guru Jambeshwar University of Science and Technology, Hisar. Her total Research publications are thirteen in International Conferences and Journals. She is working as a Reviewer/Committee member of various International Journals and Conferences.

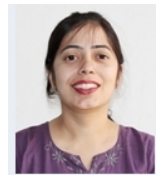